\documentclass[conference,10pt]{IEEEtran}
\usepackage{amsmath}
\usepackage{resizegather} 
\usepackage{cite}
\usepackage{graphicx}
\usepackage{fixltx2e}
\usepackage{epstopdf} 
\usepackage{array}
\usepackage{amssymb}
\usepackage[english]{babel}
\usepackage{multirow}
\usepackage{booktabs}
\usepackage{pifont}
\usepackage{indentfirst}
\usepackage{algorithm}
\usepackage{algpseudocode}
\usepackage[table]{xcolor}
\usepackage[caption=false,font=footnotesize]{subfig}
\usepackage{placeins}
\usepackage{bm}
\usepackage{array}
\usepackage{float}
\addto\captionsenglish{}
\newcommand{\norm}[1]{\left\lVert#1\right\rVert}

\newcommand{\R}{\mathbb{R}}
\usepackage{soul}

\usepackage{cleveref}
\crefname{equation}{}{}
\Crefname{equation}{Equation}{Equations} %start of sentence
\crefname{table}{Table}{Tables}
\crefname{figure}{Fig.}{Fig.}
\Crefname{figure}{Figure}{Figures}
\crefname{section}{Section}{Sections}
\usepackage{fancyhdr}
\usepackage{kantlipsum}

\fancypagestyle{plain}{
  \fancyhf{}
  \fancyhead[L]{Conference on \LaTeX}     %% C or L or R.
  \fancyfoot[L]{This is a notice}%                        %% C or L or R.

}
\usepackage{eso-pic}
\begin{document} 
\AddToShipoutPictureBG*{%
  \AtPageUpperLeft{%
    \setlength\unitlength{1in}%
    \hspace*{\dimexpr0.5\paperwidth\relax}%%  change \dimexpr0.5\paperwidth\relax appropriately
    \makebox(-4.25,-0.75)[c]{\normalsize IEEE PES General Meeting, 2016 (Accepted)}%
}}
%\AddToShipoutPictureBG*{%
%  \AtPageLowerLeft{%
%    \setlength\unitlength{1in}%
%    \hspace*{\dimexpr0.5\paperwidth\relax}%%  change \dimexpr0.5\paperwidth\relax appropriately
%    \makebox(0,0.75)[c]{\Large Notice}%
%}}
%\title{Robust Operation of Buildings in a Smart Grid Environment}
\title{Robust Reserve Capacity Provision and Peak Load Reduction from Buildings in Smart Grids}
%\title{Robust Reserve Capacity Provision and Peak Load Reduction from Utility Driven Demand-side}
%%Conference Authors
\author{
\IEEEauthorblockA{Sarmad Hanif$^{1,2}$\IEEEauthorrefmark{1}, 
	Dante Fernando Recalde Melo$^{1}$\IEEEauthorrefmark{2}, Mehdi Maasoumy$^{4}$\IEEEauthorrefmark{3}, Tobias Massier$^{1}$\IEEEauthorrefmark{6},  Thomas Hamacher$^{3}$\IEEEauthorrefmark{4}, Thomas Reindl$^{2}$\IEEEauthorrefmark{5}}

\IEEEauthorblockA{$^{1}$TUM CREATE Limited, Singapore 138602}

\IEEEauthorblockA{$^{2}$Solar Energy Research Institute of Singapore (SERIS), National University of Singapore (NUS), Singapore 117574}

\IEEEauthorblockA{$^{3}$Technical University of Munich (TUM), Garching 85748, Germany}

\IEEEauthorblockA{$^{4}$Department of Electrical Engineering and Computer Science, University of California, Berkeley, CA 94720, USA}

\IEEEauthorblockA{\IEEEauthorrefmark{1}sarmad.hanif@tum-create.edu.sg, 
	\IEEEauthorrefmark{2}e120025@e.ntu.edu.sg, 
	\IEEEauthorrefmark{3}maasoumy@eecs.berkeley.edu,\\ 	
	\IEEEauthorrefmark{6}tobias.massier@tum-create.edu.sg,
	\IEEEauthorrefmark{4}thomas.hamacher@tum.de, \IEEEauthorrefmark{5}thomas.reindl@nus.edu.sg}
}
\maketitle

\begin{abstract}
This paper proposes a robust demand-side control algorithm in a smart grid environment for heating, ventilation and air conditioning (HVAC) systems. A robust model predictive control (RMPC) scheme in a receding horizon fashion is deployed, which optimizes electricity cost and capacity market participation of the HVAC system, while satisfying comfort and operational constraints of the building and utility, respectively. Thermal load uncertainties experienced by the HVAC system are included to perform a realistic assessment of the developed controller. The National Electricity Market of Singapore (NEMS) is used as a case study and the developed RMPC scheme is tested for various price signals and scenarios. Numerical simulation results show the effectiveness of the developed framework to be readily adopted by utilities -- interested in realizing a grid-friendly and economicaly eficient demand response (DR) strategy.
\end{abstract}


\section{Introduction}
One of the key concepts in smart grids is the interaction of users and utilities. To achieve it, bidirectional communication is promised among operators, retailers and consumers. Under such conditions, one can realize adjusting local DR strategies, based on exegenous signals from utility. This can help relieve the grid's need for a higher reserve requirements, due to the integration of highly variable renewable energy supply. With a similar philosophy, various utilities have established demand response (DR) programs, aiming to improve the overall efficiency of the grid \cite{DRprog1, DRprog2, EMADR, EMAIL}. 

Among all energy consumption sectors, buildings are considered as one of the major contributors of greenhouse gas emissions and electricity consumption. Within a building, the HVAC system consumes the largest portion of the energy. Space cooling/heating along with the thermal inertia of buildings provides an inherent flexibility in the consumption of electricity. In principle, the usage of space cooling's flexibility can provide: $(1)$ reduced building operational cost, $(2)$ ancillary service provision to the grid and, $(3)$ grid secuirity.

Significant amount of work has been done for controlling energy consumption of buildings. Recent contribution regarding price-based and direct load control was reported in \cite{Vrettosa}. The applicability of MPC to control building energy consumption was reported in \cite{Olde, YMa, Mehdi4}. The use of thermal electric loads for the provision of ancillary services, and reduction of the balancing groups' scheduled deviations were reported in \cite{Vrettosb} and \cite{Vrettosd}, respectively. Authors in \cite{Mehdi5} mentioned methods for dealing with uncertain thermal load of the HVAC system, by including them as bounded disturbances.

The aforementioned papers either deal with the market participation or the uncertainties of the HVAC system model. Furthermore, the DR planning framework capable of analyzing the interaction of the market-oriented robust control scheme, subjected to various utility pricing structures and ancillary services provision has also not yet been presented.

The contribution of this paper is twofold. First, it develops an RMPC scheme to co-optimize the energy schedule, with respect to both the energy and capacity market. Second, it assesses the applicability of the developed control scheme for minimizing the total cost of the HVAC system under an exogenous peak load constraining utility signal.

The remainder of the paper is organized as follows. Section \ref{sec:1} explains the HVAC system and building model, along with the market and uncertainty settings used for developing the RMPC scheme. The RMPC control scheme is designed in section \ref{sec:2}. In section \ref{sec:3}, simulation results are presented. Section \ref{sec:4} concludes this paper with comments on adequacy of RMPC scheme for providing flexible demand, with corporation of utility peak load reduction signal.
%%%%%%%%%%%%%%%%%%%%%%%%%%%%%%%%%%%%%%%%%%
%%%%%%%%%%%%%%%%%%%%%%%%%%%%%%%%%%%%%%%%%%
\section{Modeling and Market Environment}\label{sec:1}
%%%%%%%%%%%%%%%%%%%%%%%%%%%%%%%%%%%%%%%%%%
\subsection{Modeling Framework}

The HVAC system considered in this paper has a variable air volume (VAV) mass flow rate. This provides us the opportunity to change the variable frequency drive, to meet the energy demand of a building. The cooling/heating demand is calculated based on a thermal dynamic model given in \cite{Mehdi1}. The validation of model is given in \cite{Mehdi2}. The external and internal loads are estimated in \cite{Mehdi3}. Essentially, the model provides a nonlinear relationship of the form $\dot{x}_{t} = f\left(x_t,u_t,\hat{d_t}\right)$ between the room temperature $x_{t} \in \R^{n_d}$ and the air flow input $u_{t} \in \R^{j}$ from the HVAC, experiencing disturbance $\hat{d_t} \in \R^{n_d}$. The thermal dynamic model of a room is represented as a network of $n = i + j$ nodes. Where $i$ and $j$ represent walls and room, respectively:
%%%%%%%
\begin{subequations}\label{eq:1}
\begin{align}
	\frac{dT_{wi}}{dt} &= \frac{1}{C_{wi}}\left[\sum_{j{ \in N_{wi}}} \frac{T_j - T_{wi}}{R_{ij}} + r_i \alpha_i A_i q^{''}_{radi}\right], \\
	\frac{dT_{ri}}{dt} &= \frac{1}{C_{ri}}\left[\sum_{j_{\exists N_{ri}}} \frac{T_j - T_{ri}}{R_{ij}} + \dot{m}_{ri} c_p \left(T_{si} - T_{ri}\right) \right. \nonumber \\
  &\qquad  \left. {} + w_i \tau_{wi} A_{wi} q^{''}_{radi} + \dot{q}_{int}\right],
\end{align}
\end{subequations}
%%%%%%%
Where $T_{wi}$, $C_{wi}$, $\alpha_{i}$, $A_i$ and $c_p$ represent the temperature, thermal capacitance, absorptivity factor, area, and specific heat capacity of the room $i$, respectively. $N_{wi}$ shows the set of all neighboring nodes to $w_i$. The value of $r_i$ is equal to 0 for internal, and 1 for peripheral walls. For the $i$-th room, $T_{ri}$, $C_{ri}$ and $\dot{m}_{ri}$ show its temperature, thermal capacitance and air mass flow rate, respectively. The transmittance and area of the $i$-th window is given by $\tau_{wi}$ and $A_i$, respectively. $q^{''}_{radi}$ is the solar irradiation experienced by room $i$, and $\dot{q}_{int}$ represents the internal heat generated due to equipments, furniture and occupancy. $w_i$ shows if windows are present on the surrounding walls of the room.

It can be observed from \eqref{eq:1}, the system at hand is non-linear. For the purpose of designing a controller, linear models are desirable. In \cite{Mehdi1}, author proposed a method based on Sequential Quadratic Programming (SQP), to obtain a linearized model. After the linear system is obtained, Zero-order hold is performed to discretize it. The resultant discrete time state system is represented as:
%%%%%%%
\begin{equation}\label{eq:3}
\begin{align}
x_{k+1} &= A x_k + B u_k + E \hat{d}_k
\end{align}
\end{equation}
%%%%%%% 

Where $x_{k+1} \in \R^n$ is the temperature of all the states at step $k+1$, due to control input $u_k \in \R^j$ and disturbance $\hat{d}_k \in \R^n$, at time step $k$. To calculate the cost of consumption as a function of fan power $u_k$, $K\left(u_k\right)$ is represented in \eqref{eq:10}. Where the electricity price at time step $k$ is represented as $c_k$. And sample time to convert power to energy is taken as $ \Delta t $:
%%%%%%%
\begin{equation}\label{eq:10}
\begin{align}
K\left(u_k\right)  &=  \Delta t \ c_k \left(P_{f,u_k} + P_{c,u_k} + P_{h,u_k}\right)
\end{align}
\end{equation} 
%%%%%%%

In \eqref{eq:10}, $P_{f,u_k}$, $P_{c,u_k}$ and $P_{h,u_k}$ are power consumed by the fan, cooling and heating coil of the HVAC system, respectively (more details regarding units and dimensions are given in \cite{Mehdi1, Mehdi2, Mehdi3}). 
\subsection{Model Extension}\label{subsec:1_2}
To align the model with our objectives, two modifications are performed to the original model presented above. Firstly, to account for the uncertainties into the model, additive uncertainty $w_k$ $\in$ $\R^{n_w}$ is introduced. Where $n_w$ are the number of uncertain variables. The origin of disturbance is the in-ability to model part of the thermal behaviour of the room acurately. We consider box-constrained disturbance uncertainty with uniform distribution i.e. to be known bounded by some measure, other wise unknown.
%%%%%%%
\begin{equation}\label{eq:4}
\begin{align}
x_{k+1}^{a} &= A x_{k} + B u_{k} + E (\hat{d_{k}} + w_k), \\
\mathcal{W}_k &= \{ w: \norm{w} \leq \sigma_k \}
\end{align}
\end{equation}
%%%%%%%%
where $\mathcal{W}_k$ is the set of all possible disturbance uncertainties $(\forall k = 0, 1, \ldots, N - 1)$ and $w_k \in \mathcal{W}_k$. $N$ is the prediction horizon to be used in developing the controller in section \ref{sec:2}. The dimensions of vectors and matrices in \eqref{eq:4} follows directly from the original system. Moreover, the introduction of box-constrained disturbance variable $w_k$ in \eqref{eq:4} is done by adjusting the original disturbance vector $\hat{d_{k}}$.  

The next extension to the model is performed based on the idea suggested by authors in \cite{Vrettosb}. To enable the HVAC system for offering its reserve capacity, we have extended the state space model of \eqref{eq:4} as:
%%%%%%%
\begin{equation}\label{eq:5}
\begin{align}
x_{k+1} &= A x_{k} + B u_{k} + E (\hat{d_{k}} + w_k) + B_{r} r_{k}
\end{align}
\end{equation}
%%%%%%%

At each time step $k$, the matrix $B_{r} \in \mathbb{R}^{n \times j}$ translates the effect of the extra power in the form of reserve capacity $r_k \in \mathbb{R}^{j}$ on to the temperature of zones. Matrix $B_{r}$  (by multiplying $0$ or $1$ with original input coefficients from $B$) indicates whether the HVAC system is considering to offer some reserve capacity or not. Similar to \eqref{eq:10}, the cost $K\left(r_k\right)$ of allocating reserve capacity $r_k$ for the reserve price of $b_k$ at time step $k$ is calculated by:
%%%%%%%
\begin{equation}\label{eq:12}
\begin{align}
K\left(r_k\right)  &= \Delta t \ b_k \left(P_{f,r_k} + P_{c,r_k} + P_{h,r_k}\right)
\end{align}
\end{equation} 
%%%%%%%

For the purpose of the development of a model-oriented control scheme, the modeled system presented above, with the given initial state $x_0$, is used to predict the future states of the system as:
%%%%%%%
\begin{equation}\label{eq:9}
\begin{align}
\textbf{x}_k  &= \textbf{A} x_{0} + \textbf{B} \textbf{u}_k + \textbf{E} \mathbf{\hat{d}}_{k} + \textbf{w}_{k} + \textbf{B}_{r} \textbf{r}_{k}
\end{align}
\end{equation}
%%%%%%%

Where $\textbf{x}_k  = \left[ x_{k|k}, x_{k|k+1} \ldots ,x_{k|k+N} \right]' \in \R^{n\left(N+1\right)}$ represents the predicted states at time step $k$ along a prediction horizon $N$. The subscript ``$k|k+1$" is used to denote the prediction state at time $k$ for time $k+1$. Similar explanation is valid for other predicted state vectors $\textbf{u}_k,\textbf{r}_k \in \R^{j \left(N\right)}$, $\mathbf{\hat{d}}_{k} \in \R^{n_{d} \left(N\right)}$ and $ \textbf{w}_{k} \in \R^{nw \left(N\right)}$. The matrices $\textbf{A}$, $\textbf{B}$, $\textbf{B}_{r}$, and $\textbf{E}$ are of appropriate dimensions.
%%%%%%%%%%%%%%%%%%%%%%%%%%%%%%%%%%%%%%%%%%
\subsection{Market Environment}\label{subsec:1_3}
%%%%%%%%%%%%%%%%%%%%%%%%%%%%%%%%%%%%%%%%%%
Note that in \eqref{eq:5}, the extension considers two scenarios i.e. curtailment or not-curtailment of the HVAC's load, which means the offered capacity from the HVAC system is always kept positive. This setting is used because in this paper the market framework of NEMS is used \cite{NEMS}. These two trajectories are implemented to replicate the interruptible load (IL) program, already in place in the NEMS \cite{NEMS}. In IL, the load operator can submit its bid for each 48 half-hourly period of the day. In the case of load bid getting accepted, the load operator must then curtail its offered load \cite{EMAIL}.
%%%%%%%%%%%%%%%%%%%%%%%%%%%%%%%%%%%%%%%%%%
%\begin{figure}
%%%%%%%%%%%%%%%%%%%%%%%%%%%%%%%%%%%%%%%%%%
\begin{figure}
	\centering
  \includegraphics[width=0.90\columnwidth]{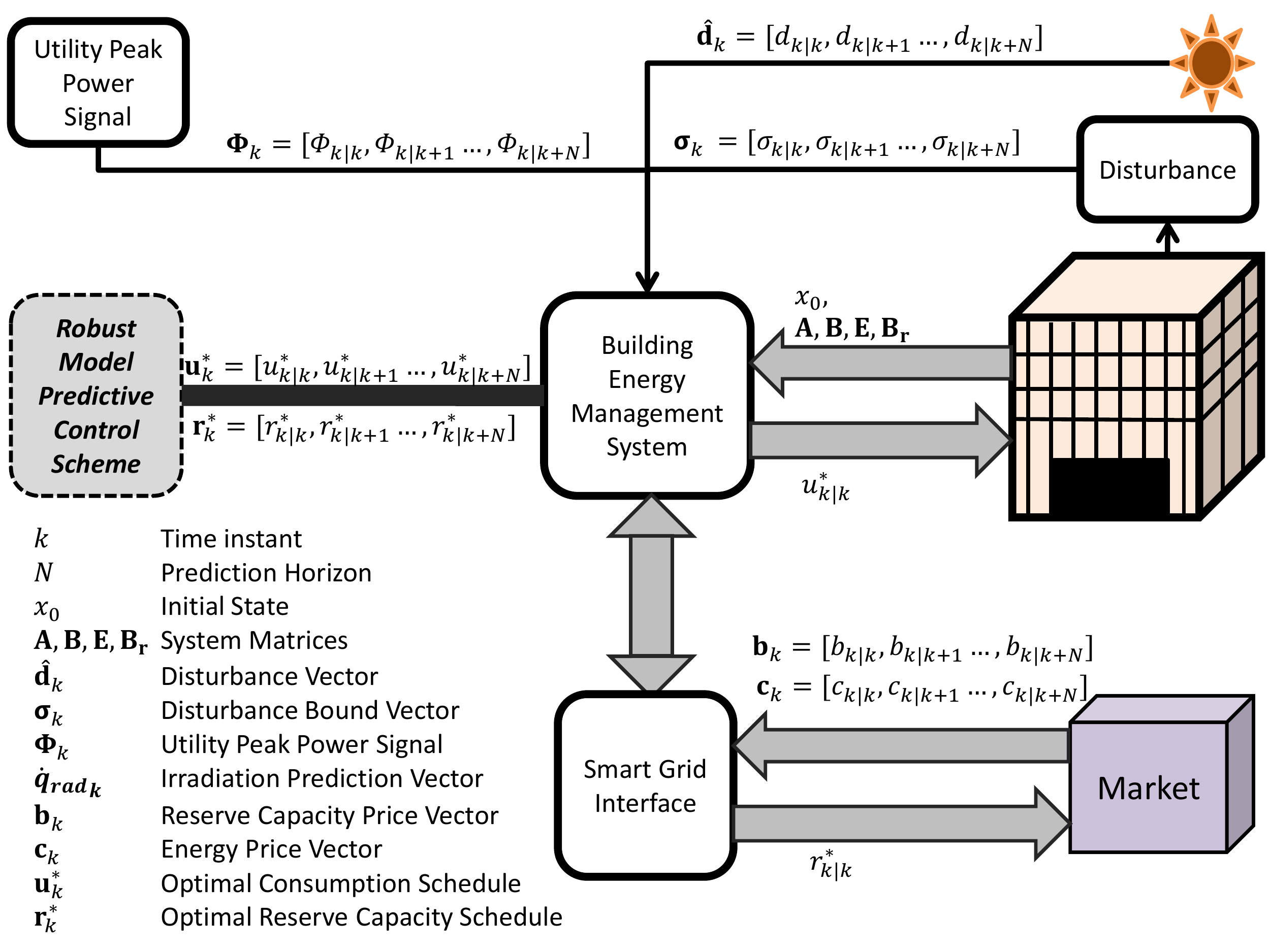}
  \caption{Experimental set-up of the RMPC control scheme.}
\label{fig1}
\end{figure}
%%%%%%%%%%%%%%%%%%%%%%%%%%%%%%%%%%%%%%%%%%
%\end{figure}
%%%%%%%%%%%%%%%%%%%%%%%%%%%%%%%%%%%%%%%%%%
%%%%%%%%%%%%%%%%%%%%%%%%%%%%%%%%%%%%%%%%%%
%%%%%%%%%%%%%%%%%%%%%%%%%%%%%%%%%%%%%%%%%%
\section{Controller Design}\label{sec:2}
As shown in \eqref{fig1}, a perfect two way communication channel between the smart grid interface (SGI) and the Building Energy Management System (BEMS) is assumed. The RMPC scheme for each time step $k$ is formulated as: 
%%%%%%%%
\begin{IEEEeqnarray}{LLr}\label{eq:11}
	\IEEEyesnumber\IEEEyessubnumber*
	\underset{\textbf{u}^{*}_{k}, -\textbf{r}^{*}_{k}}{\text{min}}  \textbf{K}\left(\textbf{u}_k\right) + \textbf{K}\left(-\textbf{r}_k\right) +  \rho \bm{\epsilon}_k + \beta_k \\
	\text{subject to} \\
		\bm{\phi}_k' \begin{bmatrix}P_{f,\textbf{u}_k} P_{c,\textbf{u}_k} P_{h,\textbf{u}_k}\end{bmatrix}  \leq \begin{bmatrix} 1 \ 1 \ 1 \end{bmatrix} \beta_k\\
	\underset{\norm{\textbf{w}_{k}} \leq \bm{\sigma}_k}{\text{max}}   \textbf{x}^{C}_{k+1} = \textbf{A}  \textbf{x}^{C}_{k} + \textbf{B} \textbf{u}_k + \textbf{E} (\mathbf{\hat{d}}_{k} + \textbf{w}_k)\label{eq:10a}\\
	\underset{\norm{\textbf{w}_{k}} \leq \bm{\sigma}_k}{\text{max}}  \textbf{x}^{NC}_{k+1}= \textbf{A} \textbf{x}^{NC}_{k} + \textbf{B} \textbf{u}_k + \textbf{E} (\mathbf{\hat{d}}_{k} + \textbf{w}_k) + \textbf{B}_{r} \textbf{r}_{k} \IEEEeqnarraynumspace \label{eq:10b}\\
	\textbf{x}{^{-}_{k}}  - \bm{\epsilon}_k  \leq \textbf{x}^{C}_{k} \leq  \textbf{x}{^{+}_{k}} + \bm{\epsilon}_k \label{eq:10f}\\
	\textbf{x}{^{-}_{k}}  - \bm{\epsilon}_k  \leq \textbf{x}^{NC}_{k} \leq  \textbf{x}{^{+}_{k}} + \bm{\epsilon}_k\label{eq:10g}\\
	\textbf{u}{^{-}_{k}} -  \textbf{r}_{k}  \leq \textbf{u}_{k} \leq \textbf{u}{^{+}_{k}} - \textbf{r}_{k}\label{eq:10h}\\
	\textbf{r}_{k}, \textbf{u}_k - \textbf{r}_{k}, \bm{\epsilon}_k\ \geq \bm{0} \label{eq:10i}
\end{IEEEeqnarray}
%%%%%%%%
 For the entire prediction horizon $N$, the solution of the optimization problem  formulated in \eqref{eq:11} results in the cost optimal schedule $\textbf{u}^{*}_{k}$ and reserve $\textbf{r}^{*}_{k}$ capacity sequence. The slack variable $\bm{\epsilon}_k$ -- penalized by a scalar $\rho$ in the objective function is implemented as a soft constraint on the upper $\textbf{x}^{+}_{k}$ and lower limits $\textbf{x}^{-}_{k}$ of both curtailed and not-curtailed scenarios. A utility peak-power-penalty (PPP) $\bm{\phi}_k$ $\in$ $\R^N$ (\$/kW) is communicated to the BEMS and the peak power term -- defined as an epigraph $\beta_k$ -- is minimized in the objective function \cite{Kelman}. The benefit of restricting the peak load in the objective function provides: $(1)$ dynamic  inclusion of an updated PPP signal from the utility signal at each time step $k$, and $(2)$ keeps the economic objective function of the RMPC scheme generic and consistent, with and without the inclusion of PPP.
 
 Constraints \eqref{eq:10a} and \eqref{eq:10b} restricts the not-curtailed $\textbf{x}^{NC}_{k+1}$ and curtailed $\textbf{x}^{C}_{k+1}$  trajectories within the feasibile region. This procedure robustify the consumption schedule of both the curtailed $(\textbf{u}_k)$ and not-curtailed $(\textbf{u}_k - \textbf{r}_{k})$ scenarios, to stay within their respective comfort zones of \eqref{eq:10f} and \eqref{eq:10g}, respectively. \eqref{eq:10h} and \eqref{eq:10i} imposes the actuator limits of the HVAC system.

The maximization term presented in \eqref{eq:10a} and \eqref{eq:10b} can be manually made robust using standard procedures presented in the literature \cite{Ben}. Essentially the procedure is to calculate the robust counterpart of the uncertain problem to yield a linear program. For the curtailed case of \eqref{eq:10a}: 
%%%%%%%%
\begin{IEEEeqnarray*}{LLr}\label{eq:12a}
\text{max \ } & \textbf{x}^{C}_{k+1} = \textbf{A}  \textbf{x}^{C}_{k} + \textbf{B} \textbf{u}_k + \textbf{E} (\mathbf{\hat{d}}_{k} + \textbf{w}_k) \\
\text{s.t. \ } &  -\bm{\sigma}_k \leq \textbf{w}_k \leq \bm{\sigma}_k \IEEEyesnumber
\end{IEEEeqnarray*}
%%%%%%%%
Using Lagrangian multipliers $\lambda_{w,1}$ and $\lambda_{w,2}$, dual of \eqref{eq:12a} can be expressed as:
%%%%%%%%
\begin{IEEEeqnarray*}{LLr}\label{eq:12b}
\text{min \ } & \textbf{x}^{C}_{k+1} = \textbf{A}  \textbf{x}^{C}_{k} + \textbf{B} \textbf{u}_k + \textbf{E} \mathbf{\hat{d}}_{k} + \bm{\sigma}_k(\lambda_{w,1} + \lambda_{w,2}) \\
\text{s.t. \ } &  \lambda_{w,1} - \lambda_{w,2} = \textbf{E} \\
&  \lambda_{w,1}, \lambda_{w,2} \geq 0 \IEEEyesnumber
\end{IEEEeqnarray*}
%%%%%%%%
When strong duality holds, then for any feasible $\lambda_{w,1/2}$ in \eqref{eq:12b}, the maximization term of \eqref{eq:12a} becomes upper bounded. Hence, the minimization term can be dropped. The resulting robust counterparts of both cases (curtailed and not-curtailed) is jointly written as:
\begin{IEEEeqnarray*}{LLr}\label{eq:13}
\textbf{x}^{C}_{k+1} = \textbf{A}  \textbf{x}^{C}_{k} + \textbf{B} \textbf{u}_k + \textbf{E} \mathbf{\hat{d}}_{k} + \bm{\sigma}_k(\lambda_{w,1} + \lambda_{w,2})\\
\textbf{x}^{NC}_{k+1}= \textbf{A} \textbf{x}^{NC}_{k} + \textbf{B} \textbf{u}_k + \textbf{E} \mathbf{\hat{d}}_{k} + \textbf{B}_{r} \textbf{r}_{k}+ \bm{\sigma}_k(\lambda_{w,1} + \lambda_{w,2}) \\
\lambda_{w,1} - \lambda_{w,2} = \textbf{E} \\
\lambda_{w,1}, \lambda_{w,2}   \geq \bm{0} \IEEEyesnumber
\end{IEEEeqnarray*}

The RMPC scheme presented above, though deployed receding horizongly, still is an open-loop control strategy. Because, while optimizing the schedule, RMPC scheme does not incorporate that the adjustment of input is also possible after the measurement of state is available for the next time step. A better approach to deal with this is the closed-loop MPC, which incorporates affine policies of the uncertainty in the optimization problem. Unfortunately, the robust counterpart of the closed-loop MPC results in larger number of variables than the open-loop MPC. Various techniques are provided in the literature on how to overcome the problem of large number of variables \cite{Lofberg}. But, for the case studies of this paper, the improvement in objective function from the closed-loop implementation was almost insignificant. Hence for the purpose of saving the computational efforts, it was decided to deploy an open-loop RMPC. In future studies. Neverthless, in the future, the sensitivity analysis with respect to the computational tractability and accuracy of the closed-loop and the open-loop MPC strategies could serve as an interesting topic.
%%%%%%%%%%%%%%%%%%%%%%%%%%%%%%%%%%%%%%%%%%
%\begin{figure}
%%%%%%%%%%%%%%%%%%%%%%%%%%%%%%%%%%%%%%%%%%
\begin{figure}[!ht]
	\centering
  \includegraphics[width=1\columnwidth]{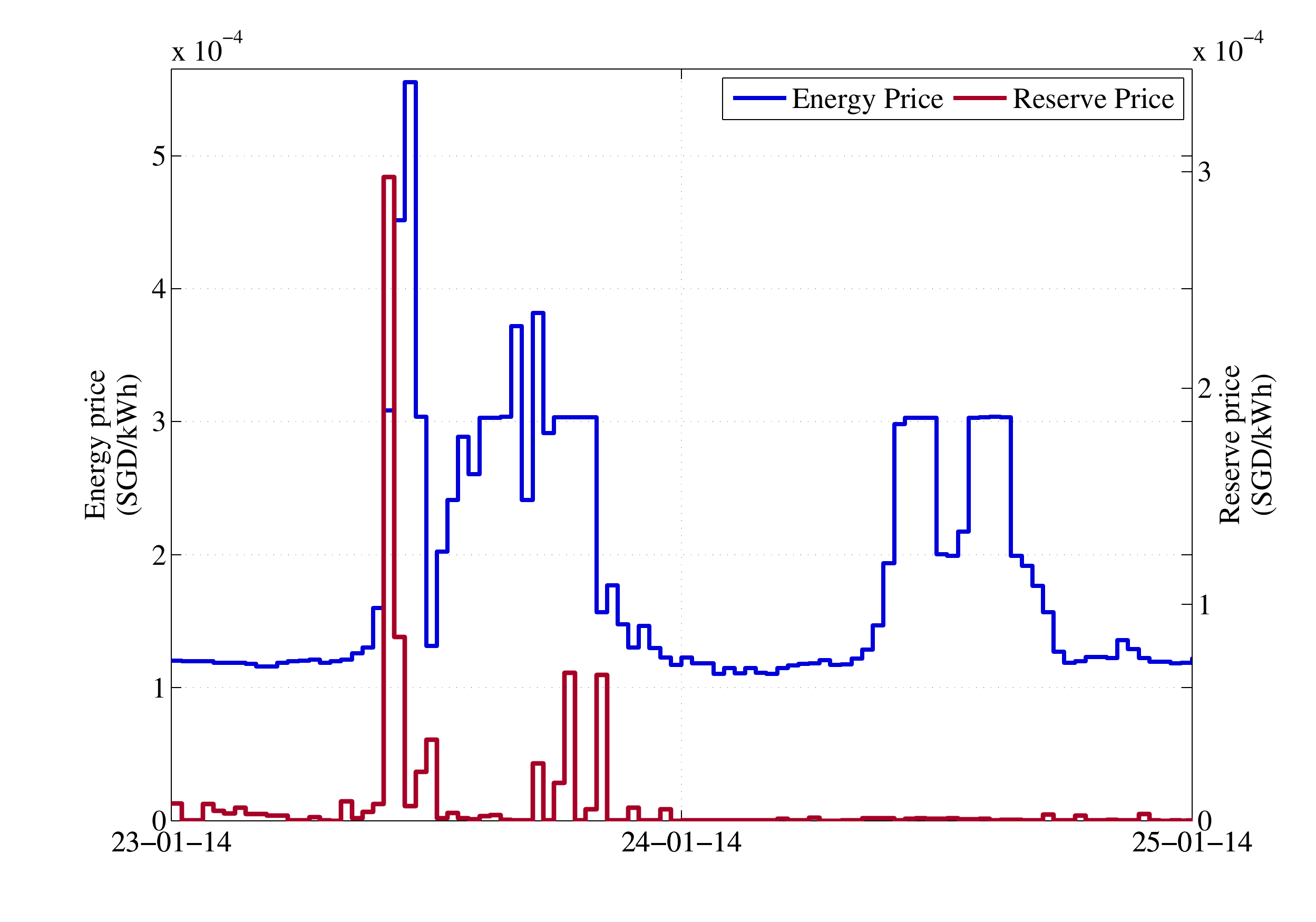}
  \caption{Real time energy and reserve price in Singaporean Dollars (SGD) taken from the NEMS.}
\label{fig2}
\end{figure}
%%%%%%%%%%%%%%%%%%%%%%%%%%%%%%%%%%%%%%%%%%
%\end{figure}
%%%%%%%%%%%%%%%%%%%%%%%%%%%%%%%%%%%%%%%%%%
\section{Simulation Results}\label{sec:3}
The linear robust optimization problem of \eqref{eq:11}-\eqref{eq:13} has been implemented using YALMIP \cite{YALMIP} and CPLEX \cite{CPLEX}. All simulations are performed with the real time energy and reserve price taken from the NEMS \eqref{fig2}. The prediction horizon of $1$ day ($48$ periods) is chosen as a compromise between the stability of the MPC and computational efforts. Since, in principle, a longer prediction horizon provides more stability to the MPC scheme. But for our simulations, prediction horizon larger than 1 day showed very little improvement in the cost, but with great increase in computational expense. $3$ Scenarios considered for evaluating the RMPC schemes are:
\begin{description}
	\item[a] Nominal MPC scheme (NMPC)
	\item[b] RMPC scheme without PPP
	\item[c] RMPC scheme with PPP of 1.5 (SGD/kW)
\end{description}
%%%%%%%%%%%%%%%%%%%%%%%%%%%%%%%%%%%%%%%%%%
%\begin{figure}
%%%%%%%%%%%%%%%%%%%%%%%%%%%%%%%%%%%%%%%%%%
\begin{figure}[!ht]
	\centering
  \includegraphics[width=0.98\columnwidth]{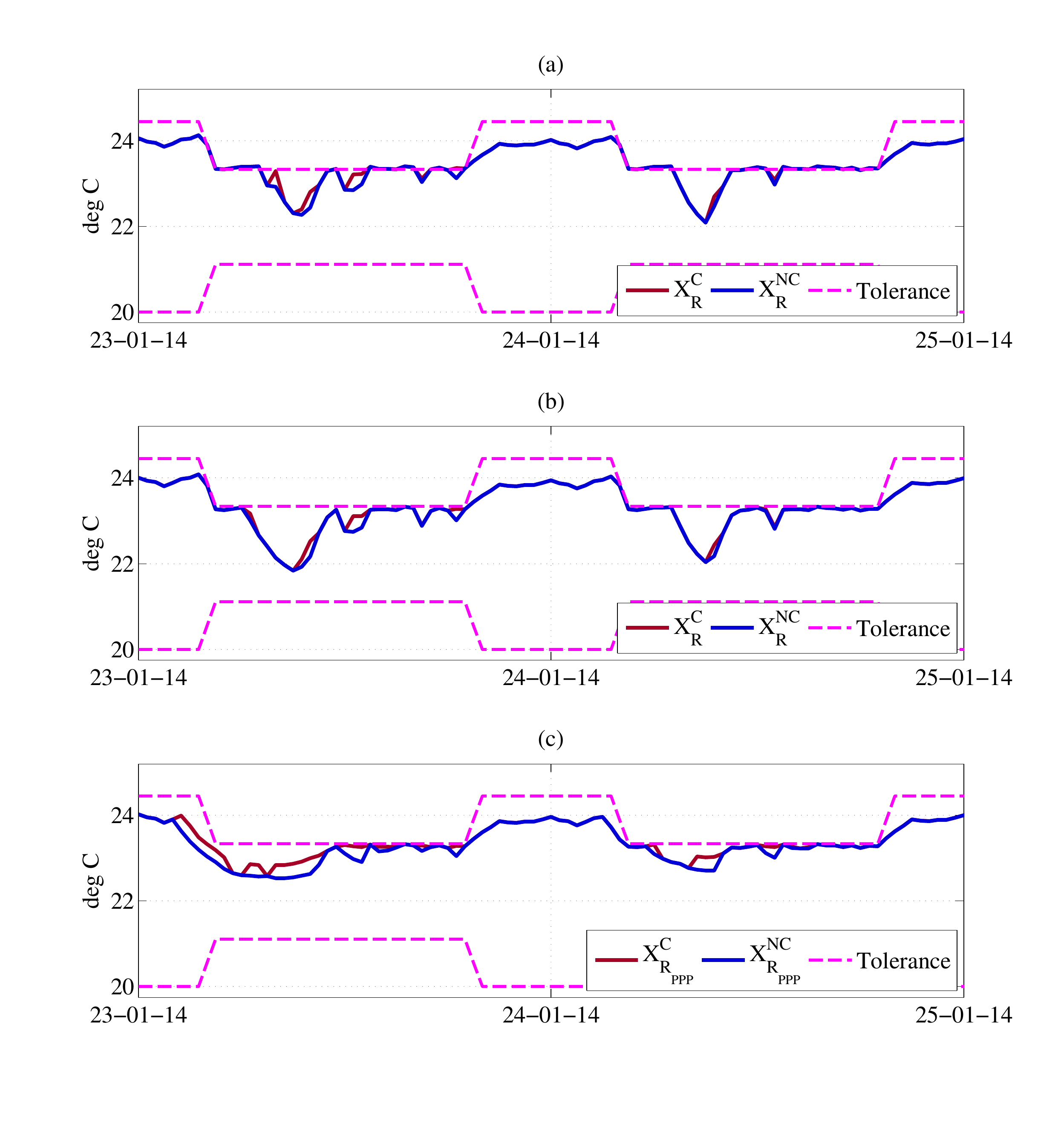}
  \caption{The curtailed and not-curtailed temperature evolution for the (a) NMPC, (b) RMPC without PPP and (c) RMPC with PPP scenarios.}
\label{fig5}
\end{figure}
%%%%%%%%%%%%%%%%%%%%%%%%%%%%%%%%%%%%%%%%%%
%\end{figure}
%%%%%%%%%%%%%%%%%%%%%%%%%%%%%%%%%%%%%%%%%%
%%%%%%%%%%%%%%%%%%%%%%%%%%%%%%%%%%%%%%%%%%
%\begin{figure}
%%%%%%%%%%%%%%%%%%%%%%%%%%%%%%%%%%%%%%%%%%
\begin{figure}[!ht]
	\centering
  \includegraphics[width=0.98\columnwidth]{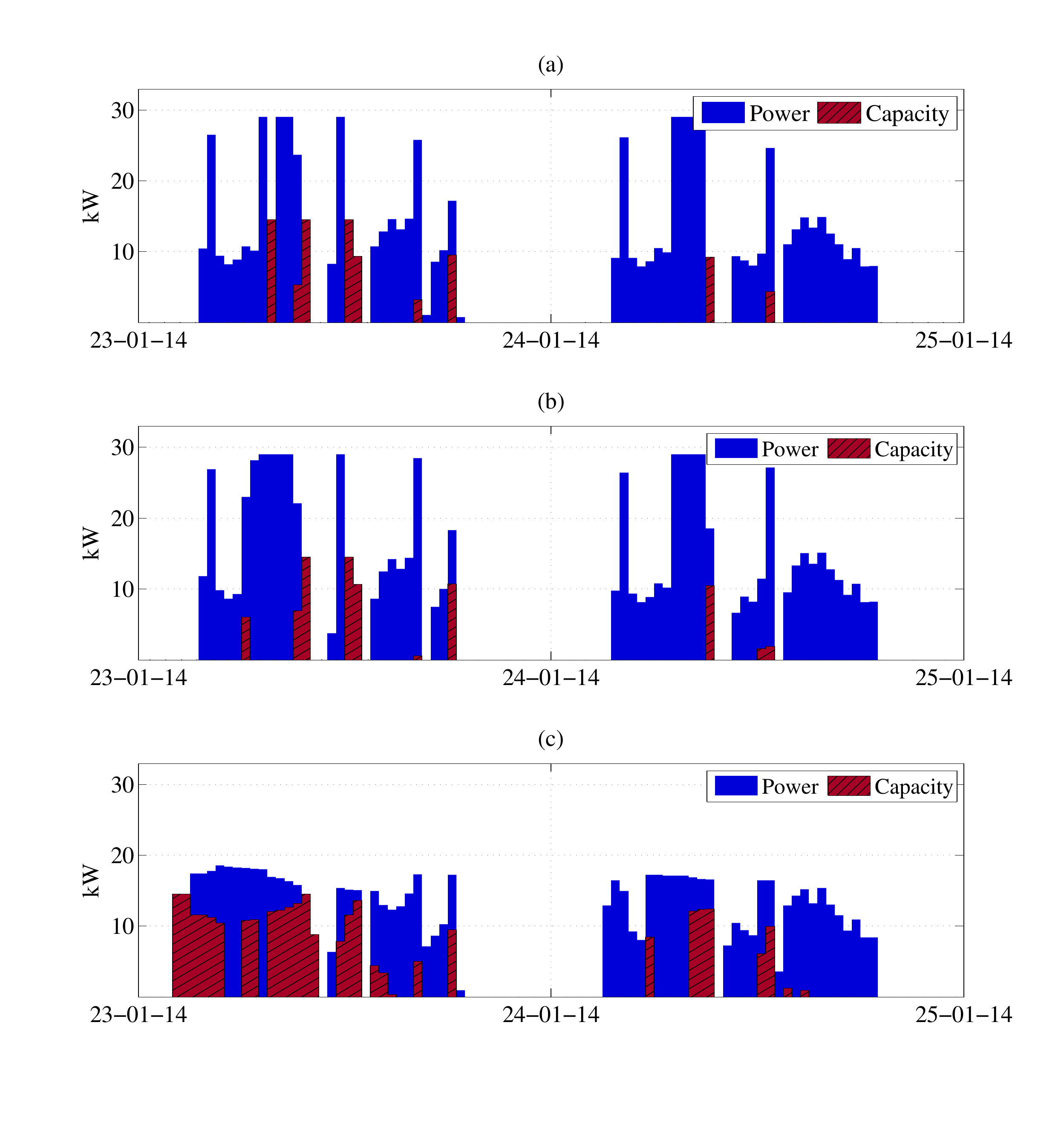}
  \caption{Power consumption and reserve capacity allocation for the (a) NMPC, (b) RMPC without PPP and (c) RMPC with PPP scenarios.}
\label{fig6}
\end{figure}
%%%%%%%%%%%%%%%%%%%%%%%%%%%%%%%%%%%%%%%%%%
%\end{figure}
%%%%%%%%%%%%%%%%%%%%%%%%%%%%%%%%%%%%%%%%%%
For all scenarios, simulation results from $2$ days of the year $2014$ are presented in fig. \ref{fig5} and \ref{fig6}. Where X{\textsuperscript{C}}{\textsubscript{*}} and X{\textsuperscript{NC}}{\textsubscript{*}} from fig. \ref{fig5} represent the curtailed and not-curtailed case for the given $*$ scenario. To evaluate the performance of the RMPC scheme, we have assumed the disturbance prediction error of approximately 50\%. That means, maximum deviation from the actual disturbance is 50\%. The NMPC scheme doesn't consider uncertainty in the system. 

Compared to the scenario in fig. \ref{fig5}(a), both RMPC schemes -- without PPP fig. \ref{fig5}(b) and with PPP fig. \ref{fig5}(c) -- demonstrate successfully that they are capable of adhering to the comfort requirements in the presence of uncertainties. One of the key observations from fig. \ref{fig5}(b) and \ref{fig6}(b) is that the RMPC scheme takes care of the uncertainty in the model at the expense of extra consumption. This is due to the fixed bounds on disturbances, which is seen by the controller as an extra thermal load to be cooled off in the room. And as a result, the controller ends up consuming some energy also at high price periods. This effect is even more pronounced in fig. \ref{fig6}(c), due to the inclusion of PPP signal. But nevertheless, the actual goal of the utility -- minimizing the overall peak load -- is achieved.

The simulation-setup of fig. \ref{sec:2} is repeated for the whole year of $2014$.  Table \ref{table1} shows monthly average cost of consumption and revenues from placing reserve capacity for the year $2014$. It can be observed that despite the increase in the cost of consumption, the increase in revenue has also occured for the scenario (b) and (c). This is due to some of the load scheduled at high price periods, providing opportunity to also allocate reserves. But due to low reserve prices, the magnitude of earnings from reserves are not comparable to the total cost of operation.

\begin{figure}[!ht]
	\centering
  \includegraphics[width=1\columnwidth]{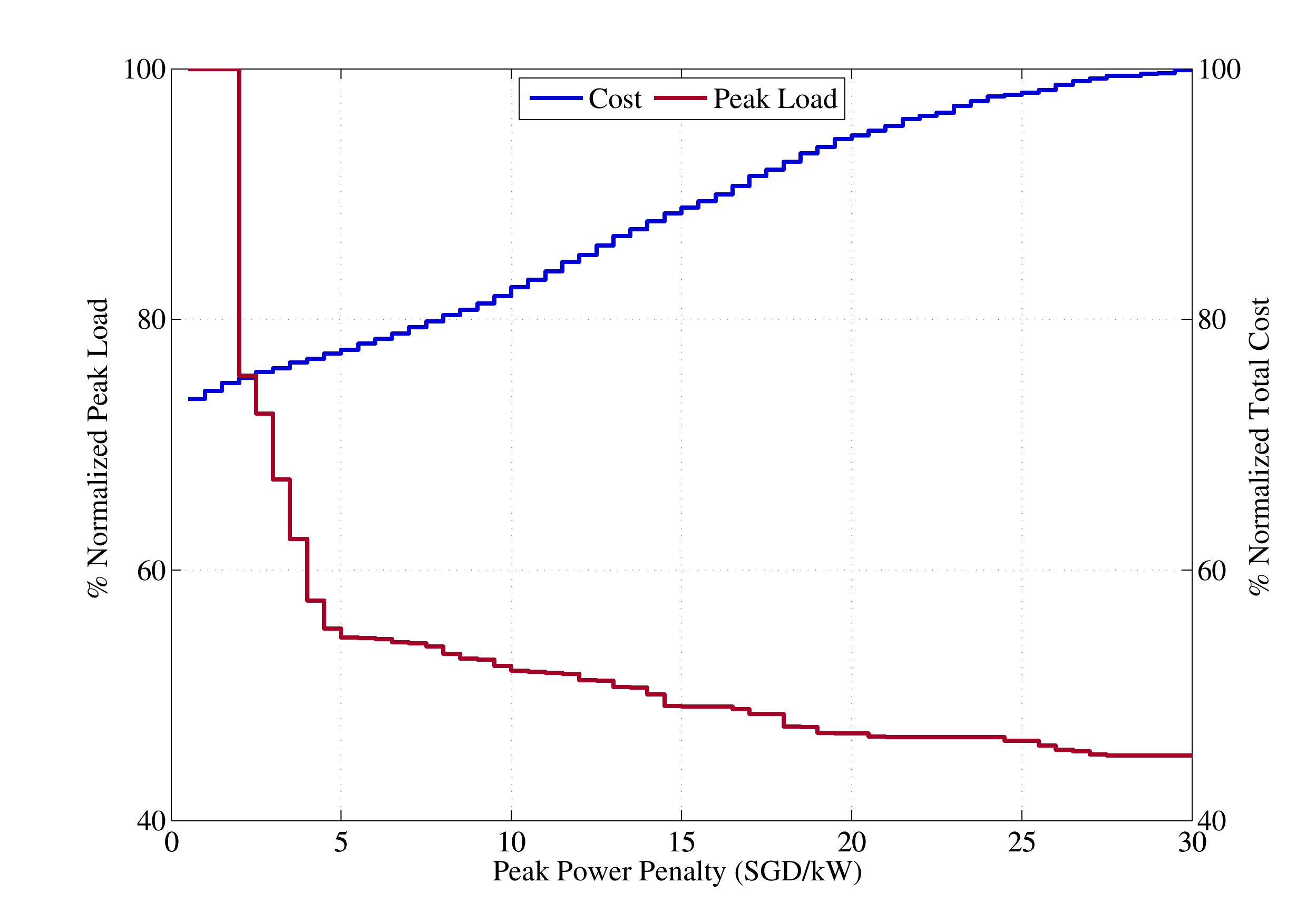}
  \caption{Reduction of peak load and as increment of total cost due to the PPP.}
\label{fig7}
\end{figure}

Fig. \ref{fig7} shows the effect of increasing PPP on the cost and the peak load reduction. Two new terms are introduced; $\%$ Normalized Total Cost $= Cost_j/Max(Cost) \ \forall j =1,2, \ldots z $ and $\%$ Normalized Peak Load $= Peak_j / Max(Peak) \ \forall j =1,2, \ldots z $ . Where $z$ values are represented by the index $j$, ranging from 0.5 SGD/kW to 30 SGD/kW.
%%%%%%%%%%%%%%%%%%%%%%%%%%%%%%%%%%%%%%%%%%
%\begin{figure}
%%%%%%%%%%%%%%%%%%%%%%%%%%%%%%%%%%%%%%%%%%
%%%%%%%%%%%%%%%%%%%%%%%%%%%%%%%%%%%%%%%%%%
%\end{figure}
%%%%%%%%%%%%%%%%%%%%%%%%%%%%%%%%%%%%%%%%%%
%%%%%%%%%%%%%%%%%%%%%%%%%%%%%%%%%%%%%%%%%%
%\begin{table}
%%%%%%%%%%%%%%%%%%%%%%%%%%%%%%%%%%%%%%%%%%
\begin{table}[!t]
\renewcommand{\arraystretch}{1.3}
\caption{Average cost per month}
\label{table1}
\centering
\begin{tabular}{llll}
\firsthline
\bfseries Scenario & \bfseries Cost (SGD) & \bfseries Revenue (SGD) & \bfseries Total Cost (SGD) \\
\hline
%& Scheduling & Scheduling & Reserves \\
(a)  & 1471 & 15.25 & 1455.8 \\
\hline
(b)  & 1530 (+4.0\%) & 15.52 (+1.7\%) & 1514.5 (+4.0\%) \\
\hline
(c)  & 1565 (+6.4\%) & 17.10 (+12\%) & 1547.9 (+6.3\%) \\
\lasthline
\end{tabular}
\end{table}
%%%%%%%%%%%%%%%%%%%%%%%%%%%%%%%%%%%%%%%%%%
%\end{figure}
%%%%%%%%%%%%%%%%%%%%%%%%%%%%%%%%%%%%%%%%%%

Fig. \ref{fig7} shows decrease in the load reduction after PPP signal of 5 SGD/kW. Whereas, the opreational cost continues to rise. Hence, the PPP beyond this value will only result in expensive operation of the HVAC system -- without providing any significant improvement in peak load reduction for the utility. 

%%%%%%%%%%%%%%%%%%%%%%%%%%%%%%%%%%%%%%%%%%
\section{Conclusion and Future Work}\label{sec:4}
%%%%%%%%%%%%%%%%%%%%%%%%%%%%%%%%%%%%%%%%%%
The results have shown that the developed RMPC scheme provides a robust control framework for the HVAC system. The developed controller optimizes energy consumption, reserve capacity provision and peak load reduction to achieve a cost effective and grid-friendly operation of buildings. It can also be seen from the presented results that to improve the overall efficiency of distribution grid; utilities and buildings must co-optimize their underlying systems. The simulation-based analysis presented in fig. \ref{fig7}, can be use as a simplified control and planning framework, to design the incentive schemes for future load management schemes. 

Future work regarding this paper is to incorporate the distribution grid constraints in the developed RMPC scheme. It is also planned to extend building model presented in this paper to grid-oriented aggregated models.  
\section{Acknowledgment}\label{sec:5}
This work was financially supported by the Singapore National Research Foundation under its Campus for Research Excellence And Technological Enterprise (CREATE) programme. This work was also sponsored by National Research Foundation, Prime Minister’s Office, Singapore under its Competitive Research Programme (CRP grant NRF2011NRF-CRP003-030, Power grid stability with an increasing share of intermittent renewables (such as solar PV) in Singapore).
\FloatBarrier

\bibliographystyle{IEEEtran}
\bibliography{IEEE_PES_GM_Robust_Capacity_bib}
%\nocite{*}

\end{document}